\documentclass[12pt]{article}
\begin{document}
\newcommand{\dd}{{\rm d}}
\renewcommand{\aa}{\hat{\rm c}}
\newcommand{\DD}{{\cal D}}
\newcommand{\dbar}{\overline\nabla}
\newcommand{\bx}{{\bf x}}
\newcommand{\bk}{{\bf k}}
\def\be{\begin{equation}}
\def\ee{\end{equation}}

\vskip 1. cm
\begin{center} {\LARGE\bf  Global Defects In Theories With 
Lorentz Symmetry
Violation.  }\\
\vskip 1. cm Musongela  Lubo\footnote{E-mail muso@ictp.trieste.it}$\mbox{}^{\mbox{\footnotesize }}$,
\\ {\em  The Abdus Salam International Centre for Theoretical Physics},
\\ {\em P.O.Box 586 , 34100 Trieste, Italy }
 \\
\vskip 0.5 cm
\vskip 1 cm
\end{center}

\begin{abstract}
 We study global topological defects  in the Jacobson-Corley model which 
 breaks Lorentz symmetry and involves up to
fourth order derivatives. There is a window in the parameter space
in which no solution exists. Otherwise, different profiles are 
allowed for the same values of the parameters.  For a scale of
 Lorentz violation  much higher
than the scale of gauge symmetry breaking, the energy densities are higher, of 
the same order or smaller than in the usual case for domain walls, cosmic
strings and hedgehogs respectively. Possible cosmological
implications are suggested.

\end{abstract}

\newpage

\section{Introduction.}

The possibility that Lorentz symmetry may be violated in some ways has been
extensively analyzed in  recent years. In particular, the robustness of
the predictions of  inflation with respect to possible transplanckian effects,
has been analyzed in this context as well as the
Hawking radiation of black holes, the dark energy scenarios and the particle
reactions at LHC \cite{trinf1,trinf2,trinf3}. Recently, the possibility
that Lorentz symmetry may be violated has been analyzed in quantum electrodynamics
\cite{elec1}. It may help to evade the GZK cut off of the
ultra high energy cosmic rays \cite{ray1,ray2}. Its possible
impact on neutrino oscillation has also attracted attention 
\cite{neu2,neu3}.

It has been argued and demonstrated in some cases that quantum gravity 
leads to effective theories with Lorentz symmetry violation, like 
non commutative field theory. Some  models also  violate Lorentz symmetry, but by
changing the dispersion relations in a way which keeps locality 
\cite{jac,cam1,cam2,cam3}.
These models break boosts but keep translation and rotation symmetry untouched.
This is achieved by modifying the kinetic term, introducing a supplementary
term which involves the three dimensional Laplacian. This extra term can be
rewritten in a manifestly covariant way by introducing a unit time-like vector
which is also dynamical. Let us point the fact that theories  like
k-inflation or k-essence have modified kinetic terms although
Lorentz symmetry is kept valid.  

For most models, particle physics interactions have been studied, especially
the dependence of the thresholds on the new scale. One knows that in
the standard model and the grand unification theories, the sector of
classical solutions is important.
Topological defects have been analyzed in contexts where
Lorentz symmetry is violated like non commutative theories 
\cite{sol1,sol2,sol3,sol4,sol5,sol6,sol7}.

In this letter we briefly analyze how solitons are affected by 
a breakdown of the Lorentz symmetry but within a theory which is still 
local. For
simplicity we focus on global defects and restrict ourselves to the 
Jacobson-Corley model. This work is organized as follows. The second section is devoted to a brief survey of
the status of solitons in field theory on a non commutative space; it will
serve as a basis for comparison. In the third section we briefly introduce
the Jacobson-Corley dispersion relation and the Lagrangian associated to it.
Adding a Higgs potential, we work out the field equation for a domain
wall and show that one has more than one profile interpolating between
the two vacuua. The fourth and fifth sections repeat the same analysis for
cosmic strings and hedgehogs. We study the energy densities of these
configurations and extrapolate our results to draw potential cosmological
implications.

\section{Solitons in non commutative theories}

Let us consider the non commutative field theory in $2+1$ 
dimensions defined by the commutation relation
\be
[x,y] = i \theta  \quad .
\ee
The energy functional is the two dimensional integral
\be
E = \frac{1}{g^2} \int d^2 z ( \partial_z \phi \star \partial_{\bar{z}} \phi
+ \theta V(\phi)) \quad ,
\ee
where the star product, in terms of a rescaled complex 
variable, reads
\be
(A \star B)(z,\bar z) = 
 \left[ \exp{\left(\frac{1}{2} ( \partial_z \partial_{\bar z'} -
 \partial_{z'} \partial_{\bar z} )\right)} A(z,\bar z) B(z',\bar{z'})
 \right]_{z'=z}  \quad .
\ee
The complex variable $z$ parameterizes the complex plane: $z=x+i y$. 

When studying solutions with rotational symmetry and finite 
energy, it is simpler to consider big values of 
$\theta$; one can then  neglect the kinetic term. The field 
equations read, in this limit,
\be
 \frac{ \partial V }{ \partial \phi} = m^2  \phi + b_3 \phi \star \phi
 + b_4 \phi \star \phi \star \phi = 0 
\ee
In the commutative case the  only the configurations 
satisfying this equation are given by $\phi= \lambda_i$ where
the $\lambda_i$ are minima of $V$. However, when
non commutativity sets in, one obtains an infinite number of 
solutions:
\be
\phi= \sum c_n \phi_n(r^2) \quad , \quad
\phi_n(r^2) = 2 (-1)^n e^{-r^2} L_n(r^2) \quad .
\ee
The $L_n$ are Laguerre polynomials while the
$c_n$ are real numbers chosen among the extrema $\lambda_i$
of the potential. The energy densities of these
solutions are proportional to the scale of non commutativity:

\be 
E \sim \frac{2 \pi \theta}{g^2} \sum_{n=0}^{\infty}
V(c_n)
\ee

Some important remarks can be drawn from these formulas. First,
the number of solutions is infinite in this model; this is linked 
to the fact that the Lagrangian of this model contains an infinite
number of derivatives. The second one is that solutions exist
even in the limiting, unphysical case where the non commutative
scale is high. Finally, the energy density is proportional to
that scale. We will basically analyze how these characteristics
are present or not in the Jacobson-Corley model.

\section{Domain walls.}

The  Jacobson-Corley dispersion relation was initially used to study
possible transplanckian imprints on the Hawking radiation. It reads
\be
 \omega^2 = k^2 + \mu k^4  \quad .
\ee
The parameter $\mu$ sets the scale where the violation of Lorentz symmetry
sets in. As we wish to avoid a  cut-off  on momenta(
or avoid imaginary frequencies), we will take $\mu$ to
be positive. At small momenta, one has the usual mass-energy relation.

The Lagrangian which leads to domain wall solutions is constructed with a
real  scalar field and possesses $Z_2$ symmetry. Putting together the
kinetic term which leads to the Jacobson-Corley dispersion relation with a 
Higgs potential, we have
\be
 {\cal L} = \frac{1}{2} \eta^{\rho \tau} 
 \partial_\rho \phi \partial_\tau \phi
 - \frac{\mu}{2}  (\Delta \phi)^2 - \frac{\lambda}{4} (\phi^2 - v^2)^2
\ee
The three dimensional Laplacian acting on the field $\Delta \phi$ breaks the 
symmetry under boosts;
it can be rewritten as a four dimensional operator by introducing a
unit vector \cite{aether}.

We shall use the dimensionless length variable $x$ defined by
$x= z/\sqrt{\mu}$. The Higgs field will be parameterized as 
$\phi= v f(x)$. Its dependence on the space coordinate is dictated by the following 
differential equation ($f^{(n)}(x)$ is the derivative of order $n$ of the function 
$f(x)$):
\be
f^{(4)}(x) - f^{(2)}(x) + \alpha f(x) (f^2(x)-1) = 0  \quad {\rm where} 
 \quad \alpha = \lambda \mu v^2 
\ee 
is the dimensionless parameter giving the square of the ratio of the two
masses: the one linked to the gauge scale by the one related to the Lorentz
breaking scale. 
We will assume that the violation of Lorentz symmetry takes place at a  very
high energy scale, much higher than the vacuum expectation value of the Higgs
field. Thus, the parameter $\alpha$ will be tiny.

The boundary conditions are $f(-\infty)=-1$ and $f(\infty)=1$. Lets us analyze the 
behavior of the field in the asymptotic region. This is done by writing, in 
the region $x \rightarrow \infty$, the decomposition $f(x)=1+ g(x)$. The function
$g(x)$ has to vanish in this limit. It obeys the differential equation 
\be
\label{eq10}
 g^{(4)}(x) - g^{(2)}(x) + 2 \alpha g(x) = 0 
\ee
which solution can be written as
\begin{eqnarray}
\label{eq11}
g(x) &=& \sum_{k=1}^{4} C_k \exp{(\beta_k x)} \quad {\rm with} \nonumber\\
\beta_1 &=& - \beta_2 = \frac{1}{\sqrt{2}} \sqrt{1-\sqrt{1-8 \alpha}} \quad , 
\nonumber\\ 
\beta_3 &=& - \beta_4 = \frac{1}{\sqrt{2}} \sqrt{1+\sqrt{1-8 \alpha}} \quad .
\end{eqnarray}

As explained above we are interested in small values of $\alpha$.
The general solution can therefore be written as
\begin{eqnarray}
\label{eq12}
g(x) &=& C_1 \exp{(\sqrt{2 \alpha} \, x)} + 
C_2 \exp{(-\sqrt{2 \alpha} \, x)} + C_3 \exp{((1-\alpha)  x)} \nonumber\\
&+& 
C_4 \exp{(-(1-\alpha) x)} 
\end{eqnarray}
so that one has two normalizable solutions, corresponding to $C_1=C_3=0$. 
At the other spatial infinity($x\rightarrow -\infty$), writing
$f(x)=-1+ g(x)$, one ends up with the same differential equation displayed
in Eq.(\ref{eq10}). Now, the normalizable modes correspond to $C_2=C_4=0$
in Eq.(\ref{eq12}).

One sees that contrary to the orthodox model which leads to a second order
differential equation, one will have here more than one 
configuration obeying the same boundary conditions:
\begin{eqnarray}
f(x) & \rightarrow & - 1 + C_1 \,  \exp{(\sqrt{2 \alpha} \, x)}   \, {\rm as} \,
 x \, \rightarrow - \infty \, {\rm and} \nonumber\\
 f(x) & \rightarrow &  1 + C_2 \, \exp{(-\sqrt{2 \alpha} \, x)}  \, {\rm as} \,
 x \, \rightarrow  \infty
\end{eqnarray}
is  one  of them
\begin{eqnarray}
f(x) & \rightarrow & - 1 + C_3 \,  \exp{((1-\alpha)  x)}  \, {\rm as} \,
 x \, \rightarrow - \infty \, {\rm and} \nonumber\\
 f(x) & \rightarrow & 1 + C_4 \,  \exp{(-(1-\alpha)  x)}  \, {\rm as} \,
 x \, \rightarrow  \infty
\end{eqnarray}
is  another. The symmetry under parity   imposes
the equalities $C_2=-C_1$ and $C_4=-C_3$.
In the first solution, the field  goes to its asymptotic behavior much slower 
than the
second one; it has a bigger spread. The fact that the two configurations 
have vanishing Higgs fields at the origin implies that the one which
goes to the boundary value slower has a sharper slope at the origin; the 
interval on which one has to integrate is smaller. Essentially, these two
behaviors conspire to give comparable total energies to the two configurations.

The density of energy per unit length is
\be
 \sigma = \int dz T_0^0 \sim \kappa(\alpha) \frac{v^2}{\sqrt{\mu}} \quad ,
\ee
in contrast to the case where Lorentz symmetry is present for which the formula
\be
\sigma \sim \sqrt{\lambda} v^3
\ee
holds. The constant $\kappa(\alpha)$ also depends on the behavior at infinity
which is chosen.
If, as it is customary, one takes the scale at which Lorentz violation takes
place to be  the Planck one, and the vacuum expectation value of the Higgs field 
to be at the GUT scale($10^{16}$ GeV), one obtains essentially that
$\kappa(\alpha)\sim 0.1$ so that the domain wall we obtain here are two orders of
magnitude heavier than in the unmodified theory. Since domains walls dominate
the energy density of the universe after a time corresponding to
\be
t_d = \frac{1}{8 \pi G \sigma} \quad ,
\ee
their take over will be much quicker in our model, in the absence of inflation.

Let us finally remark that if the scale at which the Lorentz violation 
takes place is close
enough to the scale of gauge symmetry breaking($\alpha \geq 1/8$), the quantities 
$\beta_k$ which control the behavior of the field at infinity become  
complex. For example, for $\alpha=1$ one has 
$\beta_1= 0.813442 - 0.813135 i 
\quad {\rm and} \quad \beta_2 = 0.978318 + 
  0.676097 i $.
As no combination of the solutions is real, this means there is no domain 
wall solution. This is pretty different for the orthodox theory. It means that in
the Corley-Jacobson model we are analyzing, if the scale of gauge symmetry 
breaking is sufficiently close to the one of Lorentz violation, no domain wall 
solution exists. This is a safer situation in the sense that these objects 
are a problem from the cosmological point of view; inflation is not necessary
in this setting in order to get rid of them. However, inflation will still be
needed to generate the initial density inhomogeneities.

Let us finally emphasize another important point related to the sign of the 
parameter $\mu$. The energy  of the configuration reads
\be
E = \int dz \left[ \frac{1}{2} 
 (\partial_z \phi )^2
 + \frac{\mu}{2}  (\partial_z^2 \phi)^2 + 
 \frac{\lambda}{4} (\phi^2 - v^2)^2 \right] \quad .
\ee 
For a positive $\mu$, this integrand is a sum of three positive quantities;
the integral converges only if all of them go to zero at spatial infinity; this
means as usual that the field must be in the minimum of the potential in that
region. On the contrary, when $\mu$ is negative, the second term has a negative
sign, contrary to the two others, so that it becomes possible for the integrand
to vanish in the far region without the field being in the minimum of the
potential there. The obvious choice $\phi(\pm \infty)= \pm v$
is allowed but it is not the only one.

  Let us now come to the numerical treatment. In the orthodox theory, the
equation of the domain wall profile is of second order. This means that
two initial conditions are necessary to fix unambiguously a solution. The 
vanishing of the field at the origin(due to symmetry arguments) leaves the
first derivative as a parameter which is then chosen(by the shooting method for
example) to satisfy the boundary conditions. In our case, the equation is of
fourth order. The value of the field on the wall is fixed like in the usual
case; one then chooses the first and second derivatives and fixes the third one
to attain the prescribed value at infinity. For example, taking $\alpha=0.1$,
and $f^{'}(0)= 0.233806, f^{(2)}(0)= 0, f^{(3)}(0)=-0.0214249$, one finds a 
solution which attains its asymptotic value at $x=10$.

\section{Cosmic Strings.}

Let us now turn to the  topological defect having one more dimension, i.e
the cosmic string. The global string will be a solution of the system
driven by the Lagrangian
\be
{\cal L} = \eta^{\rho \tau} (\partial_\rho \Phi) (\partial_\tau \Phi)^+ + \mu  (\Delta \Phi)   (\Delta \Phi)^+
- \frac{1}{2} 
\lambda (\Phi \Phi^+ - v^2)^2 
\ee
which displays a $U(1)$ symmetry. The Ansatz has the form 
\be \Phi= v f(r/\sqrt{\mu}) \exp{(i \theta)}  \quad .
\ee 
 Using the dimensionless
length $x$ and the ratio between the scales defined as
$\alpha$  in the preceding section, the differential equation to be solved is
\begin{eqnarray}
f^{(4)}(x) &+& \frac{2}{x} f^{(3)}(x) - \left(1 +\frac{3}{x^2} \right) f^{(2)}(x)
+ \left( \frac{3}{x^3} - \frac{1}{x} \right) f^{'}(x) + \alpha f^3(x)
\nonumber\\
&+& \left(- \frac{3}{x^4} + \frac{1}{x^2} -\alpha \right)  f(x) = 0 \quad .
\end{eqnarray}

Let us now see what happens at the boundaries. At the origin, the fields must
vanish in order for the configuration to be regular. Writing
$f(x) \sim  x^t$, one obtains the value
\be
 t= 3 \quad .
\ee
 At infinity one has to be in the vacuum so that the
energy density vanishes there. The counterpart of Eq.(\ref{eq10}) reads
\be
2 \alpha g(x) - \frac{g^{'}(x)}{x} - g^{''}(x)
+ 2 \frac{g^{(3)}(x)}{x} + g^{(4)}(x) = 0  \quad . 
\ee
and its most general non singular solution is
\be
 g(x) = C_2 \exp{(\beta_2 x)} +  C_4 \exp{(\beta_4 x)}
\ee
where $\beta_2,\beta_4$ have been given in Eq.(\ref{eq11}); the terms we 
neglected are effectively small.
So, in principle one can build  different solutions; they begin in
similar ways near the origin but go to their value at infinity with
different rates.

Concerning the energy density, one has now to integrate on the plane
orthogonal to the string. This gives an extra factor $\sqrt{\mu}$
so that the energy density per unit length reads
\be
E = 2 \pi v^2  \kappa(\alpha)
\ee 
where $\kappa(\alpha)$ is a dimensionless function. We have put a cut-off
at a radius where the field attains $95$ percent of its asymptotic value. If  gauge
symmetry is broken at the GUT scale while Lorentz 
symmetry breaking takes place at the Planck scale, this number is of
order of a few hundreds. Basically, the difference with the usual case is
not as important as for domain walls. This means such defects will behave
as the ones of the orthodox theory.

\section{Hedgehogs.}

The  Lagrangian
\be
{\cal L} = \eta^{\rho \tau} (\partial_\rho \Phi^a) (\partial_\tau \Phi_a) 
+ \mu  (\Delta \Phi^a)   (\Delta \Phi_a)
- \frac{1}{2} 
\lambda (\Phi^a \Phi_a - v^2)^2 \quad ,
\ee
where the sum on the index $a$ goes from $1$ to $3$ is $SO(3)$ symmetric.
The Ansatz for the global monopole is, in spherical coordinates,
\begin{eqnarray}
\Phi_1 &=& v f(r/\sqrt{\mu}) \sin{\theta} \cos{\phi} \quad , \quad
\Phi_2 = v f(r/\sqrt{\mu}) \sin{\theta} \sin{\phi} \quad , \nonumber\\
\Phi_3 &=& v f(r/\sqrt{\mu}) \cos{\theta}  \quad . 
\end{eqnarray}
  The profile function obeys the differential equation
\be
 f^{(4)}(x) + \frac{4}{x} f^{(3)}(x) - 
 \left(1+ \frac{4}{x^2}  \right)  f^{(2)}(x) -  \frac{2}{x} f^{'}(x)
 + \left( \frac{2}{x^2} - \alpha \right) f(x) + \alpha f^3(x) = 0 \quad .
\ee
In the asymptotic region,  one has
\begin{eqnarray}
f(x) &=& 1 + \frac{1}{x} \left( C_2 \exp{(\beta_2 x )}  +
 C_4 \exp{(\beta_4 x )} \right) \quad .
\end{eqnarray}
Note the difference with the previous defects embodied by 
the extra $1/x$ factor.
Near the origin one has  $f(x) \sim x^3$, just as for the cosmic string.

Now, the energy density has to be integrated over the three dimensional
space; this results in the formula
\be
E = 4 \pi v^2 \sqrt{\mu} \kappa(\alpha) \quad ;
\ee
from this one sees that monopoles are much lighter in this model.

Extrapolating this, it is likely that local monopoles will be much lighter
in this model; on dimensional grounds one can argue that introducing a
gauge field brings in a coupling constant. This will be
verified provided that the dependence of the 
energy on this constant is mild, as in the usual case.

\section{Conclusion.}

We have analyzed topological defects in the Jacobson-Corley model. Like in the
non commutative case, the solutions are not unique. However, they 
display  a different dependence on the scale at which the Lorentz
symmetry is broken. The hedgehogs are the only ones which energy
densities are proportional to that scale, like in non commutative theories.
On the other hand, domain walls are much heavier. As  monopoles are lighter
in this model, they will display a smaller deficit angle. The situation for
cosmic strings will be roughly like in the usual case.

This work can be extended in two ways. First, one may write down the full
equations for local defects. This would  lead to configurations
with a perfectly  integrable energy density. However, the equations are
much more involved while the main characteristics are likely to be the same.
A second point is the study of finite temperature field theory, which would
reveal the details of the formation of these defects. In the parameter space
where no defect exists, it is important to know if there is a restoration
of gauge symmetry. If this was the case, the Kibble mechanism would take 
place and the question would then be to know how the formed configurations
disappear as the universe cools.

Let us finally point out that in $ k$  inflation, the kinetic term is also
of an order higher than two, but without symmetry violation. A treatment
similar to our may be of interest in that setting.

\end{document}